# Evolution to 200G Passive Optical Network

Mamadou Diallo Diouf[1], Ahmed D. Kora[1], Octave Ringar[1,2] and C. Aupetit-Berthelemot[2]

*1. Ecole Supérieure Multinationale des Télécommunications (ESMT), BP 10000, Dakar, Senegal*

*2. XLIM UMR 6172, Dpt. C2S2, University of Limoges/CNRS, Parc Ester Technopole, 16 Rue Atlantis, BP 6804, F87068 Limoges, France*

**Abstract:** New generation passive optical network aims at providing more than 100 Gb/s capacity. Thanks to recent progress enabling a variety of optical transceivers up to 40 Gb/s, many evolution possibilities to 200G PONs (passive optical network) could be investigated. This work proposes two directly deployable cases of evolution to 200G PON based on the combination of these improved optical transceivers and WDM (wavelength division multiplexing). The physical layer of the optical network has been simulated with OptiSystem software to show the communication links performances behavior when considering key components parameters in order to achieve good network design for a given area. The complexity of the proposed architectures and financial cost comparisons are also discussed.

## 1. Introduction

Fiber Optic is the potential transmission media for next generation wired access network because of its high bandwidth capability. The number of devices and sensors providing diverse information is increasing exponentially [1]. This leads to an incredible amount of data rate which do not match as well today's access network capacity. Beside this, an existing large variety of services as internet, music on demand, gaming through an expanding set of applications (Voip, Video, email, TV, telemedicine, …) need to be improved with more quality. Many generations of access network technologies based on fiber has been deployed [2]. The first step of introduction of fiber in the access network has been to overcome the limitations of cooper in transport segment. This is known as FTTCab (Fiber to the Cabinet). One single mode fiber replaces a huge number of cooper lines. The second step has been to offer more data rate to the end users by replacing all the cooper cables by fiber optics in the other segments. The corresponding networks technologies are FTTC (Fiber to the Curb), FTTB (Fiber to the Building), FTTH/O (Fiber to the Home or Office). The FTTx (Fiber to the x) technologies different from FTTH presents major limitations and could not offer up to 100 Gps to end users. In order to optimize the deployment of fiber in the access network, passive optical network (PON) has been introduced. It is pointed out to be the most cost effective FTTH solution. A passive optical network is composed of three mains modules interconnected by fibers. These modules are the OLT (optical line termination) at central office, the splitter in the cabinet and the optical network terminal at the user side. Successive PONs technologies providing increasing access network capacity have been already deployed. The first PON technology were based on ATM (Asynchronous Transfer Mode) and denoted by APON (ATM PON). The more recent deployed and standardized passive optical network was 10G PON. The deployment of

---

**Corresponding author:** Ahmed D. Kora, Ph.D., professor, research fields: signal processing, propagation models, wireless and optical communication systems architectures and networks. E-mail: ahmed.kora@esmt.sn.

40G PON and 100G PON is expected for the next five years.

A great amount of research has been conducted lately on next generation passive optical networks. However most of these initiatives have investigated on 40G PON, 100G PON architectural problems, protocols and services [3-6]. This article presents two possible directly deployable architectures in the real world and their performances evaluated with optiwave simulation tool named OptiSystem. It takes advantages on the combination of wavelength division multiplexing and recent optical transceivers enabling up to 40 Gbps.

The rest of this article is organized as follows: We present the description of the proposed systems architectures in section 2. The section 3 is dedicated to the simulations results and discussions. Finally, we end this work with the conclusions in section 4.

## 2. Architectural System Description

The proposed network architecture is depicted in Fig. 1.

Like all passive optical network, the proposed PON architecture is composed of one OLT, one splitter and 32 or 64 ONT (optical network terminals) modules. The point-to-multipoint connectivity between the optical line terminal and ONT is ensured by the splitter. A single fiber is used to link the OLT to the splitter. From the splitter, each user is connected by its dedicated single fiber. User's area is divided in k subareas. With this configuration, the ONT modules need to be equipped with an appropriate filter according to the wavelength allocated to the corresponding subarea. Based on TDMA (time division multiples access) as access technology, a group of Nu users in the same subarea of the access network are considered to share a common bandwidth at a given allocated wavelength. The fiber and the splitter are not assumed to be lossless. In the upstream, multiple wavelength signals from allocated users systems are multiplexed at the splitter and transmitted on the common shared fiber section. From the OLT, users' data are split to all destinations but filters are used to only extract the desired wavelength at the entrance of the subareas.

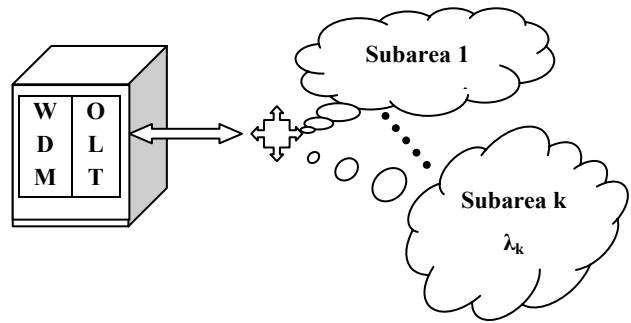

**Fig. 1   Hybrid PON network architecture.**

The PON network architecture of Fig. 1 could be approach by many potential system architectures. In this work, we have just proposed two. The first one is shown in Fig. 2. The optical transceivers of this system have a data rate of 40 Gb/s. They are supposed to transmit through the WDM (wavelength division multiplexing) module at one among the five DWDM (Dense WDM) selected adjacent wavelength in downstream (1550 nm, 1551.6 nm, 1553.2 nm, 1554.8 nm and 1556.4 nm) and five other wavelength in upstream. The WDM Mux has five input according to the described wavelength. Its bandwidth has been set at 7.23 nm and the considered noise threshold is -10 dB. The gap between the emitting band and the receiving band might minimizes waves mixing effect in the opposite transmission way. In the downstream, the OLT is supposed to send TDM (time division multiplexing) data to the end users which receivers might have already received according filtered signal based on their area filter type. In the upstream, the subscriber allocated the same wavelength share the bandwidth of 40 GHz based on TDMA (time division multiple access). The second proposed system architecture which is depicted in Fig. 3 is similar to the previous but with 20 Gb/s optical transceivers transmitting through a WDM Mux. For both systems, the optical transmitting power at the OLT has been set at 13 dBm. The modulation type is off. The extinction ratio is 10 dB and the line width is 10 MHz on off keying. The considered optical fiber could have an attenuation of 0.2 dB/Km, a 16.75 ps/nm/Km of chromatic dispersion and 0.5 ps/sqrt(Km) for PMD (polarization mode dispersion).

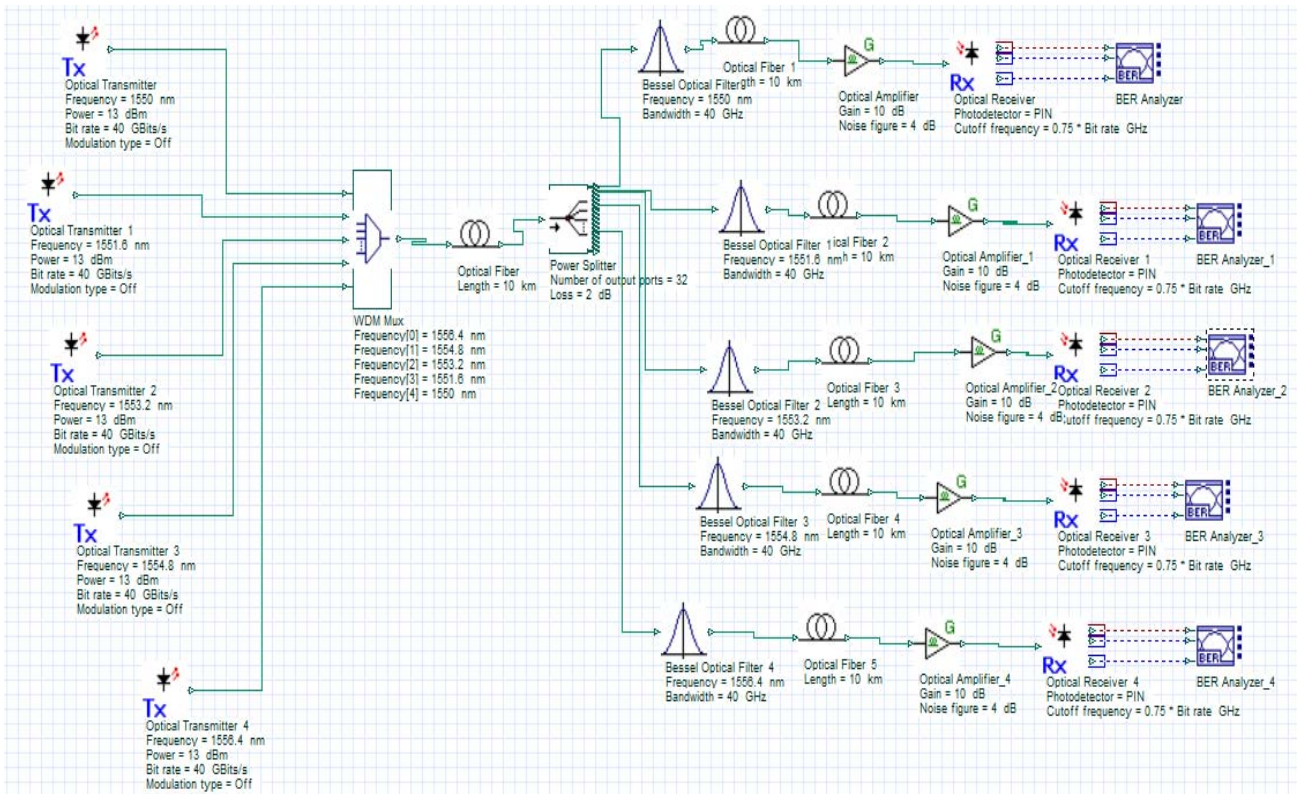

Fig. 2    200G PON based on WDM multiplexing of optical 40 Gb/s transceivers.

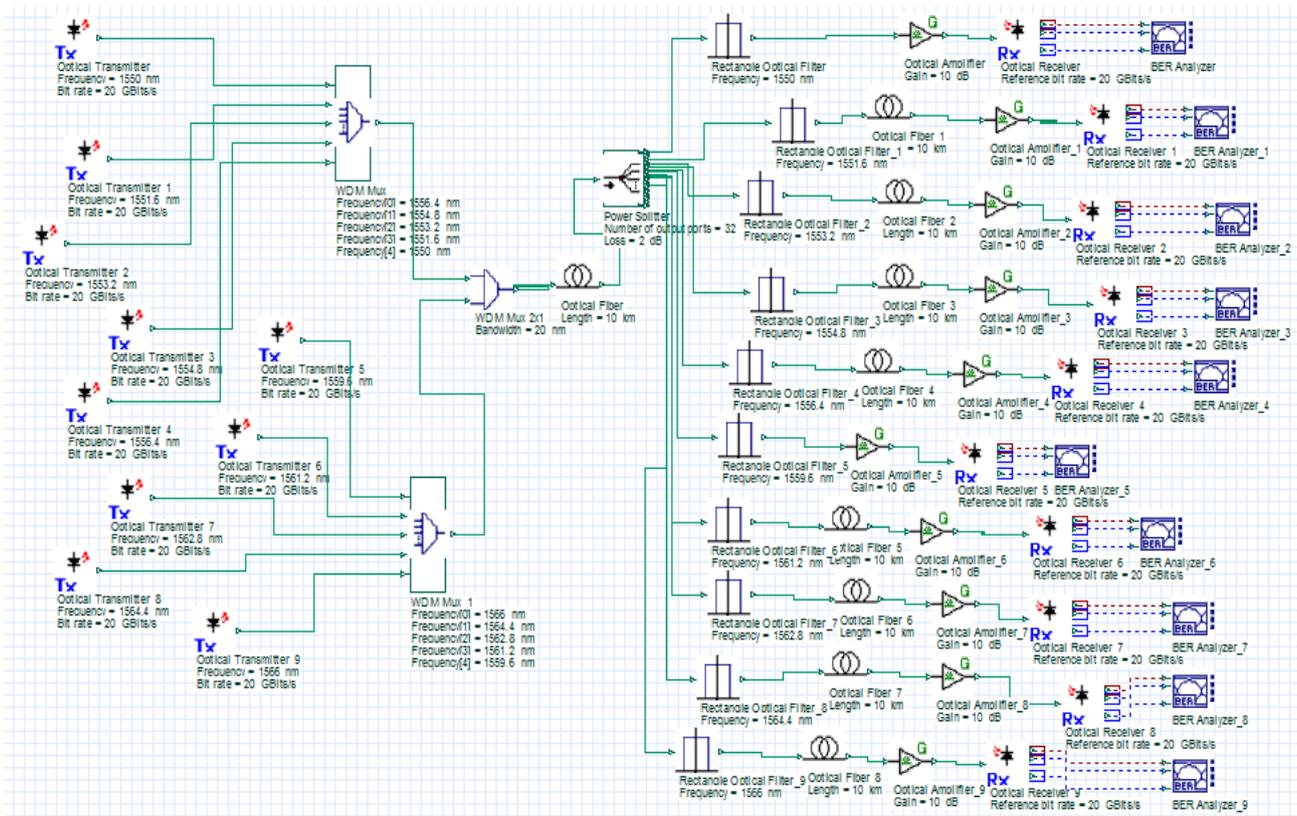

Fig. 3    200G PON based on WDM multiplexing of optical 20 Gb/s transceivers.

## 3. Results and Discussion

In order to evaluate the performance of the systems shown schematically in Figs. 2-3, a number of simulations have been performed with OptiSystem simulation tool. The initial parameters are those presented in section 2. In addition to these, the splitter number of output has been set at 32 with 2 dB of extra lost. At the receiver side, an optical amplifier set at 10 dB is associated to the optical receiver composed of PIN photo detector, a low pass filter set with a cutoff frequency 0.75 times the transmission bit rate, an approximated sensitivity of -18 dBm, a reference extinction ratio of 10 dB. The respontivity and the dark current have been set at 1 A/W and 10 nA respectively. The total distance between the OLT and the ONT is 20 Km, this includes the distance of 10 Km between the OLT and the splitter.

Figs. 4-5 are snapshots of the spectrum analyzer at one output of the splitter in the case of the first system architecture and second system architecture respectively described in section 2. These pictures show the different transmitted wavelength powers. It can also be easily observed interfering signals at the desired and other wavelengths. It is highly due to multiple waves mixing products which are potential source in this specific wavelength multiplexing. As it can be seen later, this has affected in different way the performance of the DWDM channels.

Figs. 6-7 show the impact of WDM bandwidth on the system performance. We have plotted in these figures the minimum bit error rate (min BER) for increasing bandwidth setting of WDM Mux (1.8 nm, 3.62 nm, 7.23 nm, 10 nm). When considering first order Bessel filters, it can be deduced from Fig. 6 that increasing the bandwidth of WDM Mux globally

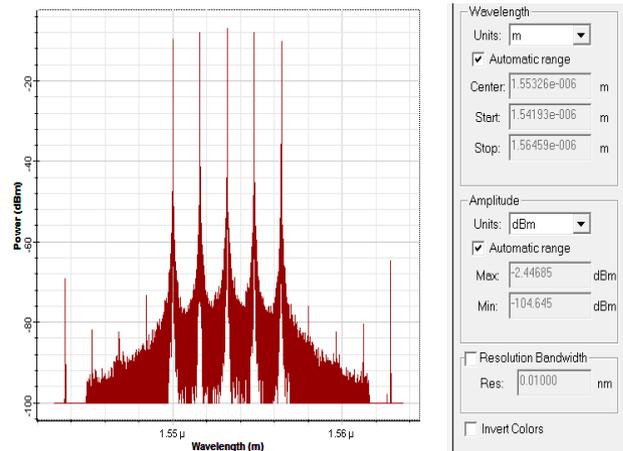

**Fig. 4 Spectrum of the multiplexed channels in the first system.**

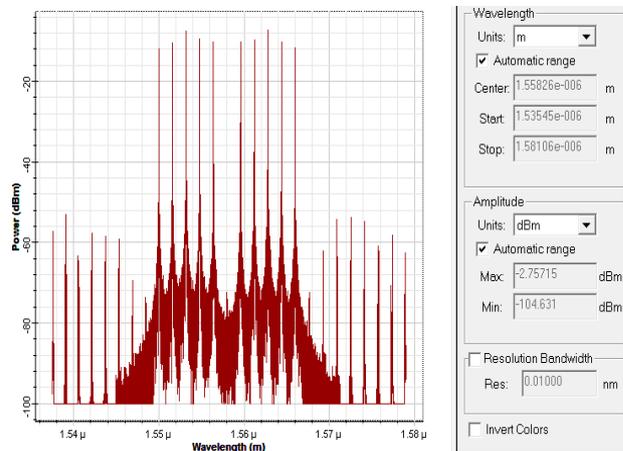

**Fig. 5 Spectrum of the multiplexed channels in the second system.**

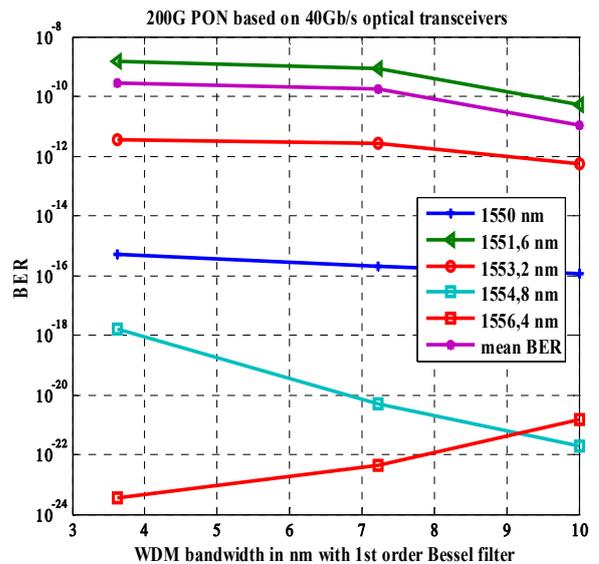

**Fig. 6 WDM bandwidth on the system performance with 1st order Bessel filter.**

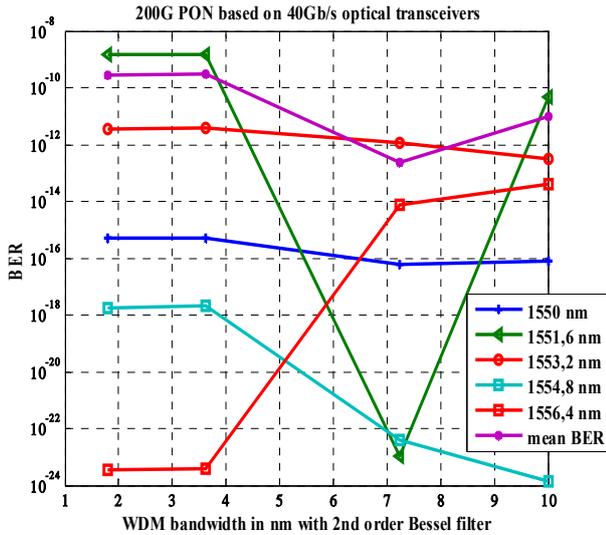

Fig. 7 **WDM bandwidth on the system performance with 2nd order Bessel filter.**

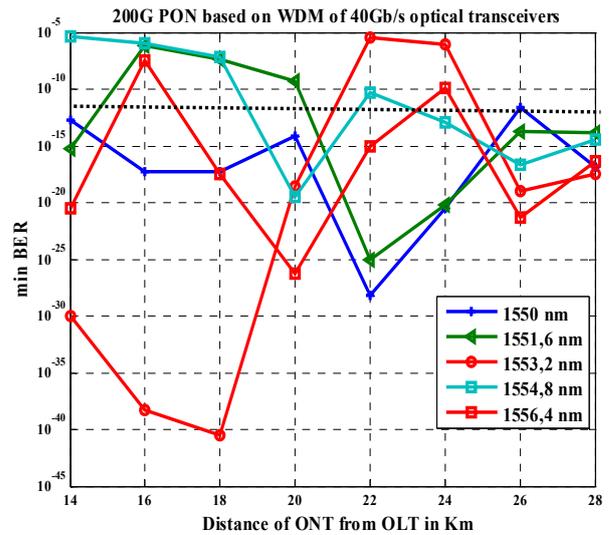

Fig. 8 **200G PON performances based on WDM of 40 Gb/s optical transceivers.**

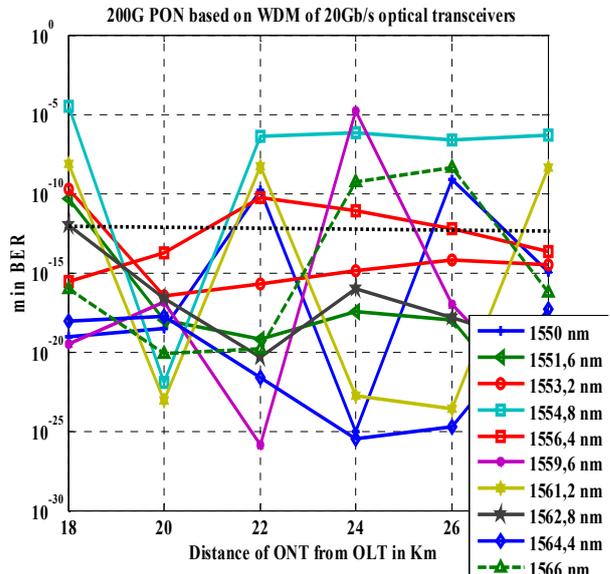

Fig. 9 **200G PON performances based on WDM of 20 Gb/s optical transceivers.**

improves the system performance but some wavelength could be exception as it is the case with 1556.4 nm. The most affected wavelengths are 1551.6 nm and 1553.2 nm with BER > $10^{-12}$.

In Fig. 7, second order Bessel filters have been considered. The simulation results show that it has improved the performance of the system in particular with the system bandwidth set at 7.23 nm where each channel presents a lower min BER (bit error rate) approximately equivalent to $10^{-12}$. A simple calculation shows that this bandwidth is closest to the required bandwidth in this case.

Since the mean min BER performance is the lowest with this bandwidth, 7.23 nm has been set wherever in rest of these simulations and the first order WDM bloc is considered.

Figs. 8-9 have pointed out the different WDM communication channel transmission quality when increasing the total distance between the OLT and ONT. The curbs describing the evolution of the min BER present a kind of sinusoidal behavior in contrast with a regular decreasing behavior. The most affected WDM channel is the centered one (1553.2 nm) with 200G PON based on 40 Gb/s optical transceivers. Fig. 8 show that the critical distance seems to be 20 Km. Up to a total distance of 16 Km, the link min BER < $10^{-30}$ and from 18 Km to 22 Km, it rapidly increases to $10^{-5}$.

The gap is more reduced in Fig. 9 with the second system based on 20 Gb/s optical transceivers where it can be seen that $10^{-25}$ < BER < $10^{-5}$. The WDM channels are also differently affected. In order to select when doing the design the most appropriate channels to allocate to subscribers in an area at a given distance up to the maximum distance limiting the area, we propose after link budget and capacity calculation to consider simulation results of the channels with BER curbs which are still under the line drawn at min BER = $10^{-12}$. For example, in Fig. 8 the WDM channel at 1550 nm is the most appropriate for all simulated total distance. It will be also more appropriate to allocate wavelengths of 1550 nm and 1553.2 nm to the users in areas up to 18 Km. The same approach could be applied to the second system with Fig. 9. In that case for example, the most appropriate WDM channels for all simulated total distance are 1562.8 nm and 1564.4 nm.

## 4. Conclusions

The evolution to 200G PON could be done in different ways. This paper has investigated two possible cases of combining WDM with TDM/TDMA based on 40 Gb/s and 20 Gb/s optical transceivers. The second system architecture is slightly more complex compared to the first one. But in term of cost of the devices, an ONT in the first case might be more expensive. Both systems need particular attention during the task of design and precisely in the selection of the wavelength to allocate to subscribers in a given area. In addition to engineering aspects, this allocation might be done according to the limitations of the considered WDM channels in term of distance and min BER.